\documentclass[12pt,a4paper]{article}
\usepackage{amsmath}
\usepackage{amsfonts}
\usepackage[dvips]{graphicx}
\usepackage{psfrag}
\usepackage[dvips]{color}
\psfrag{sf}{$S_f$}
\psfrag{hf}{$H_f$}
\psfrag{sp}{$S_p$}
\psfrag{hp}{$H_p$}
\psfrag{x}{$x$}
\psfrag{xp}{$x'$}
\psfrag{xa}{$x_A$}
\psfrag{xpa}{$x'_A$}
\psfrag{kv}{$K$}
\psfrag{kh}{Killing horizon}
\newcommand\ket{\rangle}
\newcommand\bra{\langle}
\newcommand\be{\begin{equation}}
\newcommand\ee{\end{equation}}

\begin{document}

\rightline{DCPT-04/37}   
\rightline{hep-th/0411128}   
\vfil


\begin{center} 
{\Large \bf Time-dependent spacetimes in AdS/CFT: Bubble and black hole}
\end{center} 
\vskip 1cm   
  
\renewcommand{\thefootnote}{\fnsymbol{footnote}}   
\centerline{\bf Simon 
F. Ross\footnote{S.F.Ross@durham.ac.uk} and Georgina
Titchener\footnote{G.L.Titchener@durham.ac.uk}}     
\vskip .5cm   
\centerline{ \it Centre for Particle Theory, Department of  
Mathematical Sciences}   
\centerline{\it University of Durham, South Road, Durham DH1 3LE, U.K.}   
  
\setcounter{footnote}{0}   
\renewcommand{\thefootnote}{\arabic{footnote}}


\begin{abstract}  
We extend the study of time-dependent backgrounds in the AdS/CFT
correspondence by examining the relation between bulk and boundary for
the smooth `bubble of nothing' solution and for the locally AdS black
hole which has the same asymptotic geometry. These solutions are
asymptotically locally AdS, with a conformal boundary conformal to de
Sitter space cross a circle. We study the cosmological horizons and
relate their thermodynamics in the bulk and boundary. We consider the
$\alpha$-vacuum ambiguity associated with the de Sitter space, and
find that only the Euclidean vacuum is well-defined on the black hole
solution. We argue that this selects the Euclidean vacuum as the
preferred state in the dual strongly coupled CFT.
\end{abstract}

\titlepage
\section{Introduction}

The study of time-dependent backgrounds in string theory is a key area
of development of the theory. An understanding of dynamical spacetimes
is essential for many applications of current interest, such as
cosmological evolution or black hole evaporation. Consideration of
more general spacetime backgrounds can also illuminate new aspects of
string theory, just as the consideration of quantum field theory on
more general spacetime backgrounds brought to light new effects such as
Unruh and Hawking radiation, and offered a new perspective on what the
essential elements of quantum field theory are. 

The Anti-de Sitter/Conformal field theory (AdS/CFT)
correspondence~\cite{maldads,ah} is a promising approach to the
understanding of dynamical spacetime in string theory, since the dual
field theory description is fully non-perturbative, offering a
description which in principle encompasses both the dynamics of the
background spacetime and the behaviour of strings or fields
propagating in this dynamical background. However, our understanding
of the correspondence for dynamical spacetimes remains very
patchy. The aim in this paper is to extend this dictionary by
exploring aspects of the relation between bulk and boundary for two
interesting time-dependent asymptotically locally AdS solutions.

The first solution we are interested in is the `bubble of nothing'
solution in anti-de Sitter space. This solution was introduced
in~\cite{birmingham,ross1}, following the analysis of~\cite{eva} of 
asymptotically flat bubble of nothing solutions as time-dependent
backgrounds in string theory. These solutions are constructed by the
double analytic continuation of black hole solutions, and describe a
spacetime with a compact circle direction which shrinks to zero size
on a surface which expands exponentially in the non-compact
directions. This kind of solution was originally introduced
in~\cite{wittenkk} to describe an instability of the Kaluza-Klein
vacuum. The new observation of~\cite{eva} is that they also provide
nice examples of time-dependent backgrounds for string theory, as they
are smooth vacuum solutions, which exhibit many of the issues we would
like to address in the study of time-dependence, such as particle
creation and non-trivial vacuum ambiguities for quantum
fields. In~\cite{birmingham,ross1}, an asymptotically locally AdS
bubble was constructed by double analytic continuation of the
Schwarzschild-AdS black hole solution (similar solutions were studied in~\cite{cvetic}). In~\cite{ross1}, the
counterterm subtraction procedure was used to obtain the stress tensor
of the dual field theory.

At large distances, this `bubble of nothing' solution approaches a
locally AdS spacetime. This locally AdS spacetime is in fact just a
quotient of AdS, and was interpreted previously in~\cite{ban,bgm} as a
black hole solution. That is, this is the higher-dimensional analogue
of the BTZ black hole~\cite{btz,bhtz}. This is a time-dependent
solution: there is no Killing vector which is timelike everywhere
outside the event horizon. Although it is not as smooth as the bubble
of nothing solution, this is clearly an interesting example of a
time-dependent geometry in its own right, and we will see that it has
some interesting properties.

These two solutions have the same asymptotics, so they should be
related to different states in the same field theory on the asymptotic
boundary. This boundary is a de Sitter space cross a circle: the
circle corresponds to the direction that is compactified in the
locally AdS black hole, and which degenerates at the bubble in the
bubble of nothing solution. Our aim is to explore the properties of
the spacetimes, and relate them to the dual field theory. We will
focus on understanding the relation between horizons in the spacetime and
the boundary theory, and considering the question of choices of vacuum
state in the bulk and the boundary.

In the next section, we give a brief review of the two solutions. We
then discuss the horizons in these solutions in section~\ref{hor}.  We
show that the solutions have Killing horizons which can be interpreted
as cosmological or acceleration horizons respectively, and which are
in both cases naturally related to the de Sitter cosmological horizon
in the boundary geometry. In particular, we show that the entropy of
the bulk and boundary horizons agree if we introduce a cutoff at large
radius. This sets up a novel correspondence relating horizons in the
bulk and boundary, as opposed to relating horizons in the bulk to
thermal states in the boundary. We argue that the black hole event
horizon, on the other hand, does not have a thermodynamic
interpretation. 

Thus, the thermodynamic properties in the bulk are identified with the
thermodynamic properties of the de Sitter horizon in the boundary
theory. We are proposing that the time-dependence of the bulk
spacetime is completely encoded in the time-dependence of the boundary
spacetime, and the appropriate state in the dual CFT is simply some
natural vacuum state on this curved, time-dependent background. These
examples are thus a particularly simple context for further
investigations of time-dependence in string theory, since we have
reduced the problem to the more well-understood one of studying
quantum field theory in a time-dependent background. 

One of the most interesting features of time-dependent spacetimes is
that they do not have unique vacuum states. We would therefore like to
use our solutions as a laboratory for studying the description of
vacuum ambiguities in AdS/CFT. The natural vacuum ambiguity to
consider in this context is the $\alpha$-vacua in de Sitter space,
since the boundary has a de Sitter factor, and there are coordinates
in the bulk which write these spaces in de Sitter slicings. It has
been known for some time~\cite{chernikov,tagirov,mot,all} that de
Sitter space has a one-parameter family of vacuum states invariant
under the de Sitter isometry group, called $\alpha$-vacua. There is a
unique member of this family which has the same short-distance
singularity as in flat space~\cite{mot,aj}, which is also the
Euclidean vacuum obtained by analytic continuation from the
sphere~\cite{dr,gh}. There has nonetheless been considerable
controversy in the literature about whether the additional de
Sitter-invariant vacua are physical, particularly focusing on the
definition of an interacting theory~\cite{bros1,bros2,banks,el1,uhd1,gl1,el2,collins1,gold2,collins2,dBJM,collins3}.
It is not our intention to address any of the issues raised in this
literature; instead, we want to consider a different approach, using
the behaviour of the analogues of the $\alpha$-vacua in a free scalar
field theory in the bulk spacetime to obtain information about
$\alpha$-vacua in the strongly-coupled dual field theory.

In section~\ref{alpha}, we show that there are natural analogues of
the $\alpha$-vacuum ambiguity in these two bulk spacetimes, which we
identify with the ability to choose $\alpha$-vacua in the boundary
theory. In section~\ref{bhalpha}, we find that the propagators for the
$\alpha$-vacua on the locally AdS black hole have additional
singularities at the event horizon of the black hole. We show that
these additional singularities are reflected in a breakdown of the
procedure of~\cite{bernard} for constructing the expectation value for
the stress tensor. We also comment that the analytic continuation procedure
used in~\cite{mal1,kos} to probe the region behind the event horizon
can only be extended to the locally AdS black hole if we take the
Euclidean vacuum. We therefore argue that the analogues of
$\alpha$-vacua are not good vacuum states for the locally AdS black
hole, since they break down on the event horizon. 

We interpret this as evidence that the $\alpha$-vacua are not good
states for the strongly-coupled CFT on the boundary, which is de
Sitter space cross a circle. Although the $\alpha$-vacua appear to be
acceptable states at least at the free level for the bubble of nothing
solution, this spacetime only exists when the size of the circle is
less than a maximum value. Thus, at least for a range of parameters,
there is no obvious spacetime interpretation for the $\alpha$-vacua in
the CFT, whereas there is a spacetime interpretation for the Euclidean
vacuum. Thus, the Euclidean vacuum is selected as a preferred state in
the strongly-coupled CFT.

We summarize and present some concluding remarks in
section~\ref{concl}. 

\section{Review of bubble \& black hole solutions}

The bubble of nothing solution  is obtained by
analytic continuation of the 5d Schwarzschild-AdS black
hole\footnote{We focus on the case of AdS$_5$ for definiteness, but it
  is a simple exercise to extend our remarks to other AdS$_d$ with $d
  \geq 4$.}, 
\be 
ds^2 = -(1 + \frac{r^2 }{l^2 }-\frac{r_0^2}{r^2 } )dt^2
+(1+\frac{r^2}{l^2} - \frac{r_0^2}{r^2 })^{-1}dr^2 + r^2(d\vartheta ^2
+\cos^2\vartheta d\Omega_2^2) 
\ee
where $d\Omega_2^2 = d\theta^2 + \sin^2 \theta d\phi^2$ is the metric
on the two-sphere.  By analytically continuing two variables, $t \to
i\chi $ and $\vartheta \to i\tau$, a novel solution of gravity with
negative cosmological constant is found:
\be
ds^2=(1 +\frac{r^2 }{l^2 } - \frac{r_0^2 }{r^2 })d\chi^2 + (1
+\frac{r^2 }{l^2 } - \frac{r_0^2 }{r^2 } )^{-1}dr^2 +r^2[-d\tau^2 +
\cosh^2\tau (d\theta^2 + \sin^2 \theta d\phi^2)]. 
\label{bubble}
\ee
We see that the proper length of the spacelike $\chi$ direction goes
to zero at $r=r_+$, where $r_+$ is the root of $f(r) = l^2 r^2 + r^4 -
r_0^2 l^2$,
\be
r_+^2 = \frac{l^2}{ 2} \left[ -1 + \sqrt{1+\frac{4 r_0^2 }{l^2} } \right].
\ee
To make the spacetime smooth at $r=r_+$, the coordinate $\chi$ must be
identified periodically, with period
\be 
\Delta \chi = \frac{ 2\pi l^2 r_+ }{2 r_+^2 + l^2 }.
\label{deltachi}
\ee
There is no region of spacetime inside the surface $r=r_+$; this is
the `bubble of nothing'. Since the metric on this bubble is
three-dimensional de Sitter space with scale $r_+$, we see that the
bubble expands exponentially. At very early times, the region
`excised' is a very large sphere.  As $\tau <0 $ increases, the size
of the bubble shrinks to a minimum at $\tau=0.$ After this it grows
again exponentially.
 
As was pointed out in ~\cite{ross1}, the solution (\ref{bubble}) is
asymptotically locally AdS: that is, at large distances, it approaches
\be
ds^2 = (1+\frac{r^2}{l^2}) d\chi ^2 + (1+\frac{r^2}{l^2})^{-1}dr^2
+r^2[-d\tau ^2 +\cosh^2\tau (d\theta ^2 + \sin ^2 \theta d\phi^2)],
\label{bh}
\ee
which is a locally AdS spacetime. This spacetime is the result of
quotienting AdS by a boost isometry to make the coordinate $\chi$
periodic. This coordinate
system is related to embedding coordinates for AdS by 
\begin{equation} \label{embed1}
\begin{aligned}[m]
x^1 &=(r^2+l^2)^{1/2} \cosh \chi/l, \\
x^2 &= r \sinh \tau, \\
x^3 &= (r^2 + l^2)^{1/2} \sinh \chi/l,  \\
x^4 &= r \cosh \tau \, \sin \theta \, \sin \phi, \\
x^5 &= r \cosh \tau  \, \sin \theta \, \cos \phi,  \\  
x^6 &= r \cosh \tau \cos \theta,
\end{aligned}
\end{equation}
where $x^\mu$ are embedding coordinates in terms of which AdS is
defined by $-(x^1)^2-(x^2)^2+(x^3)^2+(x^4)^2+(x^5)^2+(x^6)^2=-l^2$. If
$\chi$ is allowed to run over all values (\ref{embed1}) provides a
coordinatization of a part of AdS. Making $\chi$ periodic with some
arbitrary period $\Delta \chi$ thus introduces discrete
identifications on AdS along a boost isometry. As is evident from the
form of the metric (\ref{bh}), the quotient preserves an $SO(1,1)
\times SO(1,3)$ subgroup of the original $SO(2,4)$ isometry
group. Note that we are free to choose any period $\Delta \chi$ we
wish in this quotient geometry.

\begin{figure}[htbp]
\begin{center}
\psfragscanon
\includegraphics[scale=0.5]{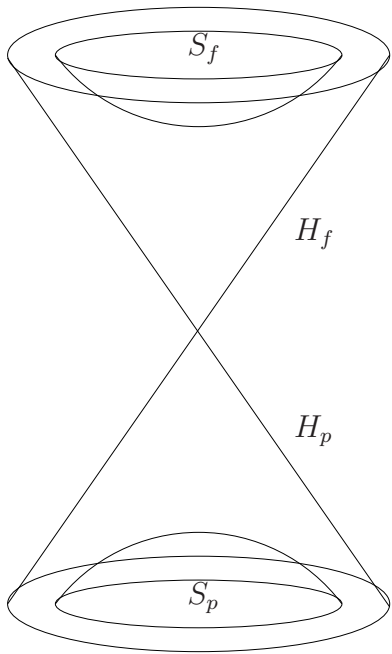}
\caption{ Three dimensions of the AdS black hole spacetime: one sphere
  direction and the $S^1$ factor are suppressed. The direction of
  increasing  $y_0$ is up. $S_f$, $S_p$ are the future and past
  singularities; $H_f$, $H_p$ are the future and past horizons. \label{fig1}}
\end{center}
\end{figure}

As was stressed in~\cite{cai}, this spacetime has been studied
previously~\cite{ban,bgm}, as the higher-dimensional analogue of the
BTZ black hole. It was also discussed in the recent classification of
quotients of anti-de Sitter spaces~\cite{ross2}. It describes an
interesting non-stationary black hole solution with a single exterior
region. The structure of this solution is more easily understood by
passing to a `Kruskal' coordinate system~\cite{bgm},
\begin{equation} \label{embedk}
\begin{aligned}
x^1 &=l {1+y^2 \over 1-y^2} \cosh \chi/l,  \\
x^2 &= 2l {y^0 \over 1-y^2}, \\
x^3 &= l {1+y^2 \over 1-y^2} \sinh \chi/l, \\
x^4 &= 2l {y^1 \over 1-y^2}, \\
x^5 &= 2l {y^2 \over 1-y^2}, \\  
x^6 &= 2l {y^3 \over 1-y^2},
\end{aligned}
\end{equation}
where we have written $y^2 = y^\mu y_\mu  =
-(y^0)^2+(y^1)^2+(y^2)^2+(y^3)^2$. In terms of these
coordinates, the radial coordinate of (\ref{bh}) is 
\be 
r= l \frac{2 \sqrt{y^2}}{1-y^2},
\ee 
and the coordinates $(\tau,\theta,\phi)$ in (\ref{bh}) parametrise the
hyperboloids $y^2 =$ constant. The metric in these Kruskal
coordinates is 
\be
ds^2 =   {4 l^2 \over (1-y^2)^2}  (-dy_0 ^2 + dy_1 ^2 + dy_2 ^2 +dy_3 ^2
) + {(1+y^2)^2 \over (1-y^2)^2} d\chi^2. 
\label{bhk}
\ee
One of the great advantages of this coordinate system is that it
writes the metric at a constant $\chi$ as conformal to a flat space,
enabling us to easily picture the causal structure of the spacetime.
We see that the singularity at $r=0$ in (\ref{bh}) is just a
coordinate singularity, corresponding to the light cone of the origin,
$y^2=0$ in the new coordinates. At $y^2=-1$, the $\chi$ circle becomes
null, and beyond this surface it will be timelike, so the quotient
introduces closed timelike curves in this
region. Following~\cite{ban,bgm}, we assume this region of closed
timelike curves is removed from the spacetime. Then $y^2=-1$ becomes a
singularity, since timelike curves end on it. The surface at $y^2=0$
then becomes an event horizon for the spacetime; observers who cross
it will inevitably hit the singularity. The asymptotic boundary of the
spacetime is at $y^2=1$. The geometry is depicted in
figure~\ref{fig1}. These Kruskal coordinates cover the whole
spacetime. 

This is the natural higher-dimensional analogue of the BTZ black hole,
but it clearly has a somewhat different global structure: the
maximally extended spacetime described by (\ref{bhk}) has only a
single exterior region, with a connected asymptotic
boundary. Furthermore, the global event horizon at $y^2=0$ is not a
Killing horizon for any Killing vector, and one can easily see that
the area of its cross-sections increase with time. We will discuss the
horizons in this solution and their interpretation further in the next
section. 

To understand the relation of these asymptotically locally AdS
spacetimes to a dual field theory, we need to understand the
conformal boundary of these spacetimes. Adopting a conformal factor
\mbox{ $\Omega = l/r$,} we see that the boundary metric for (\ref{bubble}) is
\be
ds^2_\Sigma = d\chi^2 + l^2 [-d\tau^2 +
\cosh^2\tau (d\theta^2 + \sin^2 \theta d\phi^2)]. 
\label{bdy}
\ee
We can obtain the same result from (\ref{bhk}) (in
those coordinates $\Omega = (1-y^2)/2\sqrt{y^2}$).  The dual CFT
thus lives in a space which is three dimensional de Sitter space cross
a circle, dS$_3 \times S^1$. There is a single dimensionless parameter
characterising this boundary geometry, the ratio of the radius of the
circle to the size of the de Sitter factor, $\Delta \chi/l$. This is
thus the physical information we can specify from the field theory
point of view, and it will determine the bulk geometry.

For the locally AdS black hole, there is a unique bulk geometry for
each choice of $\Delta \chi$. For the bubble of nothing, on the other
hand, the value of $r_0$ characterising the bulk geometry is
determined by solving (\ref{deltachi}), which does not give a
one-to-one map between $\Delta \chi$ and $r_0$. There is a maximum
value of $\Delta \chi$ for which this equation has a solution, $\Delta
\chi_{max} = \sqrt{2} \pi l$, attained when $r_+^2 = l^2/2$, that is,
when $r_0 = l/2$. If we choose $\Delta \chi$ greater than this maximum
value, there is no corresponding bubble of nothing solution in the
bulk. If we choose $\Delta \chi$ less than the maximum, there will be
two solutions, with a smaller and a larger value for
$r_0$ {}\footnote{Note also that the $\chi$ circle is contractible in the
bulk for the bubble of nothing solution, which fixes the bulk spin
structure to be antiperiodic around this circle. Thus, this bulk
geometry will only contribute to the path integral when we take
antiperiodic boundary conditions for the fermions on this circle in
the boundary. There is no such restriction for the locally AdS black
hole solution.}. It was argued in~\cite{ross1} that the solution with
the smaller value of $r_0$ will be both classically and quantum
mechanically unstable, and should therefore be disregarded, while we
expect the other solution to be stable.

In~\cite{ross1}, the boundary stress tensor for this solution was
computed, using the counterterm subtraction
procedure~\cite{bk,hs,skend}. If we use the bubble of nothing as the
bulk solution, we obtain a boundary energy density
\be
\rho_{bubble} = - {1 \over 16 \pi G l^3}(r_0^2 + l^2/4) = -{ N^2 \over 8 \pi^2
    l^4} \left( {r_0^2 \over l^2} + {1 \over 4} \right).
\label{bubblee}
\ee
This calculation can also be applied to obtain the boundary stress
tensor when we use the locally AdS black hole as the bulk solution, by
setting $r_0=0$ in the previous result~\cite{cai}. Thus, for this case
\be
\rho_{bh} = - {1 \over 64 \pi G l} = -{ N^2 \over 32 \pi^2
    l^4}.
\label{bhe}
\ee
We note that the energy of the bubble is lower than the energy of the
black hole, so it is possible for the black hole to decay into (some
excited state on) the bubble of nothing. It was also pointed out
in~\cite{ross1} that the stress tensor obtained for the black hole
corresponds precisely to a geometrical contribution associated with
the curvature of the background. This can presumably be interpreted as
a reflection of the fact that the solution is simply a quotient of
global AdS, so it corresponds to the dual field theory in a vacuum
state where there is no state-dependent contribution to the
stress-energy.

\section{Horizons \& thermodynamics}
\label{hor}

We now consider the relation between horizons and thermodynamics.  The
relation between the bulk black hole horizon and the thermal behaviour
of the field theory on the boundary for the ordinary BTZ and
Schwarzschild-AdS black hole solutions was one of the first things to
be understood in the context of the AdS/CFT
correspondence~\cite{strom,wittenbh}.  The black hole solution
described above has a global event horizon, but as noted previously,
this is not a Killing horizon, and it is not clear if it should have a
thermodynamic interpretation.

\begin{figure}[htbp]
\begin{center}
\psfragscanon
\includegraphics[scale=0.5]{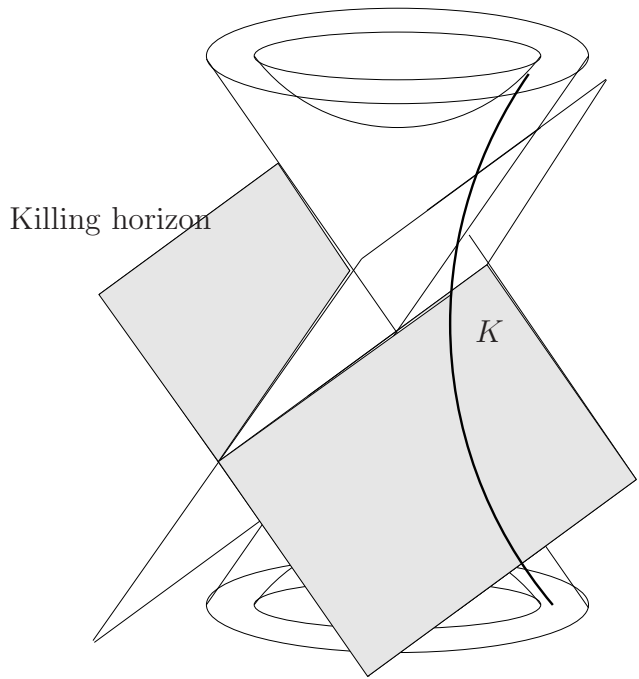}
\caption{ A timelike orbit of the Killing vector $K=\cos \theta
\partial _{\tau} +\tanh \tau \sin \theta \partial_{\theta}$ and the
plane where it becomes null, $\tanh \tau = \pm \cos \theta.$ \label{fig2} }
\end{center}
\end{figure}

Both these spacetimes do however have Killing horizons in them,
associated with the timelike Killing vectors that generate the
worldlines of comoving observers in the coordinates of
(\ref{bubble},\ref{bh}). If we consider a given comoving observer, say
the one at $\theta=0$, then the corresponding Killing vector is
\be
K = \cos \theta \partial_\tau + \tanh \tau \sin \theta \partial_\theta
\label{kvec}
\ee
and the Killing horizon where this Killing vector becomes null is at 
\be
\tanh \tau = \pm \cos \theta.
\label{chor}
\ee
Note that the location of this Killing horizon is independent of $r$
in these coordinates, and corresponds precisely to the usual Killing
cosmological horizons in the de Sitter factor. These horizons are
illustrated in figure~\ref{fig2}. 

In the bubble of nothing solutions, these Killing horizons correspond
to cosmological horizons, just as they do in de Sitter space.  It was
noted in~\cite{eva} that the bubble of nothing solution constructed
from the Schwarzschild black hole has cosmological horizons, as the
exponential expansion of the bubble prevents any one observer from
seeing the whole of the spacetime. The same is true in the AdS bubble
of nothing (\ref{bubble}). Any observer's trajectory will
asymptotically approach a constant coordinate position on the
two-sphere, because of the exponential growth of the two-sphere's
proper volume. The trajectory will thus lie within the region bounded
by the corresponding Killing horizon, and that Killing horizon will be
the cosmological horizon for this observer.

In the locally AdS black hole solution, these horizons are more
naturally identified as acceleration horizons. Any observer who
chooses to remain outside the black hole will asymptotically approach
constant values of the angular coordinates, and the corresponding Killing
horizon will be an event horizon for this observer by the same
argument. However, it is clear from (\ref{bhk}) that these observers
are accelerating--uniformly accelerating if they remain at constant
values of $r$. The Killing vector (\ref{kvec}) is also, in these
coordinates, simply a boost:
\be
K = y^0 \partial_{y^3} + y^3 \partial_{y^0}.
\ee
Thus, in the locally AdS black hole solution, we can think of these
horizons as analogous to the Rindler horizons in flat space. 

What we would like to explain now is the interpretation of these
horizons in the dual field theory. The novelty here is that the
horizons are non-compact, and intersect the asymptotic boundary. This
will imply a rather different relation between the horizons and the
dual: the horizon in the bulk is related to a horizon in the boundary,
rather than to thermal effects from considering a non-trivial mixed
state in the boundary field theory. Since the structure of this
cosmological/acceleration horizon is very similar in these two
spacetimes, we will treat them together.

First, let us consider the thermal properties of the state. Since the
horizon (\ref{chor}) is a Killing horizon, there is a unique regular
(i.e., Hadamard) vacuum invariant under the action of this Killing
vector. It will be a thermal state with respect to the notion of
time translation defined by this Killing vector~\cite{kaywald}. This
state is clearly the usual Euclidean vacuum, obtained by analytic
continuation from the Euclidean versions of
(\ref{bubble},\ref{bh}). The thermal properties of this state have a
natural interpretation from the boundary point of view: also in the
boundary, we have a cosmological horizon at (\ref{chor}), and the
natural CFT vacuum state defined by analytic continuation from the
Euclidean version of the boundary geometry will look thermal from the
point of view of comoving observers. That is, the bulk state is
identified with a vacuum state in the CFT, and looks thermal simply
because the CFT lives in a time-dependent background, a de Sitter
universe (cross a circle). 

However, there is a fly in the ointment for this very natural
interpretation: we have two bulk geometries. How do we understand the
difference between them from the boundary point of view? It seems that
the circle plays the crucial role here. In the bubble of nothing
solution, the fermions must be antiperiodic on this circle, whereas in
the locally AdS black hole, we are free to choose either spin
structure. We suggest that the bubble of nothing is related to a CFT
vacuum with antiperiodic boundary conditions on the fermions, while
the locally AdS black hole can be related to a CFT vacuum with
periodic boundary conditions on the fermions\footnote{Since it admits
both spin structures, the black hole can also contribute when we
consider antiperiodic boundary conditions. In that context, it is
presumably interpreted as an excited state above the vacuum described
by the bubble solution. It is only when we consider antiperiodic
boundary conditions that the black hole can decay into a
bubble.}. This is analogous to the identification of global AdS with
the NS ground state and the M=0 BTZ black hole with the RR ground
state in the usual AdS$_3$ story~\cite{ah}. Note however that in our
case neither state is supersymmetric, as the background dS$_3 \times
S^1$ geometry breaks all supersymmetry. As evidence in support of this
suggestion, we note that the difference in energy between the bubble
(\ref{bubblee}) and the black hole (\ref{bhe}) goes like $1/\Delta
\chi^4$ for small $\Delta \chi$, which is the expected behaviour for
the Casimir energy associated with such a change in boundary
conditions for fermions. The greatest problem with this suggestion is
that the bubble of nothing solution only exists for $\Delta \chi \leq
\Delta \chi_{max}$. We have no interpretation to suggest for this
restriction, which seems very unnatural from the CFT point of view.

Having proposed a relation between thermal properties of the states in
bulk and boundary, we would like to go on to make a more controversial
suggestion, that there should also be an entropy associated with these
horizons, by showing it has a natural interpretation in the boundary
theory. One might think that this stands little chance of working,
since the area of the bulk horizon is infinite, so the entropy $S =
A/4G$ would also be infinite. How can we give an interpretation for
this infinite entropy in terms of the boundary theory? However, this
is precisely the right answer from the boundary point of view: the
horizon in the bulk should be related to the cosmological horizon in
the boundary. This has finite area, since it is just the usual de
Sitter horizon, but the boundary theory is not coupled to gravity. We
can formally include a gravitational term with $G=0$, so the entropy
for this horizon is indeed infinite.

To make a quantitative comparison, we introduce a cut off at
$r=R$. The entropy of the bulk horizon inside this surface is
\be
S_{bulk} = \frac{A}{4G} = \frac{1}{4G}\int_{r_i}^R rdr\int d\phi d\chi
\label{bulks}
\ee
where $A$ is the area of the horizon's bifurcation surface $\tau=0,
\theta=\pi/2$, which we have written out explicitly in terms of an
integration over $(r,\phi,\chi)$. The lower limit of integration
$r_i = r_+$ in the bubble of nothing solution and $r_i=0$ in the
locally AdS black hole, and $G=G_5$ is the gravitational constant in the bulk.

In the boundary theory, the introduction of the cut off at $r=R$
corresponds to an ultraviolet cutoff on the field theory, and
introduces a coupling of the field theory to gravity, given by
calculating the induced Einstein action obtained  from integrating the bulk
action over the radial direction. The action in the bulk is
\be
I = \frac{1}{16 \pi G}\int \sqrt{-g} d^5 x R.
\ee
If we set 
\be
ds^2 = f(r) d\chi^2 + f(r)^{-1} dr^2 + {r^2 \over l^2} \widehat{ds}^2
\ee
we can rewrite this action as
\be
I = \frac{1 }{16 \pi G}\int \sqrt{-\hat{g}} \frac{r^3}{l^3}drd^4x
\frac{l^2}{r^2}\hat{R} + \ldots,
\ee
where $\hat{R}$ is the curvature of the three-dimensional metric
$\widehat{ds}^2$, and $\ldots$ denotes terms involving $f(r)$. This allows
us to read off the Newton's constant in four dimensions as
\be
\frac{1 }{G_4 } = \frac{1 }{l G}\int_{r_i}^R rdr .  
\ee 
The entropy of the bifurcation surface $\tau=0,\theta=\pi/2$ in the
boundary theory is
\be
S = {A \over 4 G_4} = {1 \over 4G_4} l \int d\phi d\chi,
\ee
which then agrees precisely with (\ref{bulks}). Thus, we see that the
entropy of the horizons in bulk and boundary agrees quantitatively. 

Thus, the AdS/CFT correspondence identifies the entropy of the bulk
cosmological/acceleration horizon with the entropy of the de Sitter
horizon in the boundary. This calculation gives a powerful argument
that even for non-compact horizons, the area should be regarded as an
entropy, as argued in~\cite{jac,jpar}. 

One final point to note concerning the cosmological horizons is that
there is a finite difference in entropy between the bubble of nothing
solution and the locally AdS black hole: the entropy of the bubble of
nothing is less than that of the black hole by 
\be
\Delta S = {2 \pi r_+^2 \Delta \chi \over 4G},
\ee
because of the different ranges of integration. This difference was
absorbed in a change in the induced Newton's constant in the boundary
from the point of view of the cutoff CFT discussed above, but it would
be interesting to see if it could be related to some difference in the
corresponding states, perhaps in the un-cutoff CFT. 

Finally, what about the global event horizon in the locally AdS black
hole? Should there be some entropy associated with this horizon as
well? We will argue that the answer is no. First, we note that there
is no independent temperature associated with this horizon. An
observer outside the black hole will see a thermal bath, but this will
come to them from the acceleration horizon that bounds the region of
spacetime they can see, and not from the event horizon, which lies
behind this acceleration horizon. Secondly, the event horizon is
not a special surface in a spacelike slice of the spacetime: it is
not, for instance, the boundary of a region of trapped
surfaces. Indeed, as noticed previously, the event horizon in this
spacetime is rather like the light cone of the origin in flat
space. It is a purely teleological event horizon, which is a boundary
of the past of infinity because of something that is going to happen
in the future: the `singularity' at $y^2 = -1$. The light cone of the
origin in flat space can similarly form part of an event horizon, if
a collapsing shell of matter is converging on it. However, we do not
think we would associate an entropy with this horizon before the shell
crossed it. 

Therefore, we think the appropriate generalisation from the
three-dimensional BTZ black hole, where the Killing horizon and event
horizon are the same thing, is to associate an entropy with the
Killing horizon in the locally AdS black hole in higher dimensions,
and not to associate any entropy with the event horizon in this solution.

\section{$\alpha$-vacua in bubble and black hole}
\label{alpha}

In the previous section, we have focused on relating the properties of
the Euclidean vacuum state on the bulk spacetimes to the dual
CFT. However, one of the most interesting features of time-dependent
spacetimes is that they do not have unique vacuum states. We would
therefore like to use our solutions as a laboratory for studying the
description of vacuum ambiguities in AdS/CFT.

The natural vacuum ambiguity to consider in this context is the
$\alpha$-vacua in de Sitter space. It has been known for some
time~\cite{chernikov,tagirov,mot,all} that de Sitter space has a
one-parameter family of vacuum states invariant under the de Sitter
isometry group. Our boundary geometry, which is de Sitter cross a
circle, will clearly inherit this ambiguity, and we would now like to
relate it to the bulk spacetime.

\subsection{Review of $\alpha$-vacua in de Sitter}
\label{alpharev}

To begin, we review the $\alpha$-vacuum ambiguity in de Sitter
space. Let $G_E$ be the Wightman function for a massive scalar field
$\phi$ on three dimensional de Sitter space obtained by analytically
continuing the unique Wightman function on the Euclidean
sphere to 3d de Sitter.  Then we define the Euclidean vacuum state
$|E\rangle$ by
\be G_E(x,x') = \langle E | \phi(x) \phi(x') |E\rangle .  
\ee 
We can choose a mode expansion for a free massive scalar field $\phi$
\be
\phi(x)= \sum_n \left( a_n \phi_n^E (x) +a_n^{\dag}\phi_n^{E*}(x) \right),
\ee
such that the vacuum state obeys
\be 
a_n |E\rangle =0.
\ee 
The Wightman function can  be re-written in terms of these modes as
\be
G_E(x, x') = \sum_n \phi _n^E (x) \phi_n^{E*}(x').
\ee
We can choose the positive frequency Euclidean modes such that 
\be
\phi_n^E(x_A)=\phi_n^{E*}(x),
\ee
where $x_A$ is the point antipodal to $x$ in de Sitter space. If $x$
has coordinates $\theta, \phi$ and $\tau$, then $x_A$ has coordinates
$\pi -\theta , \phi+\pi, -\tau$.

The $\alpha$-vacua are defined by observing that for any $\alpha \in
\mathbb{C} \text{ with Re } \alpha < 0,$ we can define a new mode
expansion by the Bogoliubov transform~\cite{mot,all}
\be \label{eq:mamodes} 
\tilde{\phi}_n(x) = N_{\alpha}(\phi_n^E(x) + e^{\alpha}
\phi_n^{E*}(x)), \quad  N_{\alpha} = \frac{1 }{ \sqrt{1- e^{\alpha
      +\alpha^*}} } .
\ee
The operators $a_n$ must also be transformed, and the new operators
are given by 
\be 
\tilde{a}_n = N_{\alpha}(a_n^E - e^{\alpha *}a_n^{E\dag} ) .  
\ee
A new de Sitter invariant vacuum state $|\alpha \rangle $ is then
defined by
\be
\tilde{a}_n | \alpha \ket  =0 \quad \forall \quad  n > 0
\ee
and the Wightman function in this vacuum is
\be
G_{\alpha} (x, x')  = \langle \alpha | \tilde{\phi}(x)
\tilde{\phi}(x') |\alpha \ket = \sum_n  \tilde{\phi}_n (x)
\tilde{\phi}_n^*(x') . 
\ee
This new propagator can be expressed in terms of the original
Euclidean modes and $\alpha$. It is \cite{bms}
\begin{multline} \label{eq:galpha}
G_{\alpha} (x, x')= N_{\alpha}^2\sum \left[ \phi_n(x)\phi_n^*(x') +
  e^{\alpha +\alpha^* }\phi_n(x')\phi_n^*(x) \right. \\ 
\left. + e^{\alpha ^* } \phi_n(x)\phi_n^*(x'_A)+e^{\alpha
}\phi_n(x_A)\phi_n^*(x') \right].
\end{multline}
As a function  of Euclidean Wightman propagators, it can be written as
\be \label{gafromge}
G_{\alpha }(x, x') = N_{\alpha}^2 \left( G_E (x, x') +e^{\alpha
  +\alpha ^*}G_E(x', x) + e^{\alpha^*}G_E(x, {x'}_A)
+e^{\alpha}G_E(x_A, x') \right). 
\ee 
These new vacua are automatically invariant under the continuous
$SO(1,3)$ symmetry of the de Sitter space, as can be seen from the
above relation between $\alpha$ and Euclidean propagators. If we take
$\alpha$ to be real, the $\alpha$-vacuum is also invariant under time
reversal, which interchanges the last two terms in \eqref{gafromge}.

Unlike the Euclidean vacuum, the $\alpha$-vacuum is not
thermal. In~\cite{burg,bms}, the departure from thermality was studied by
considering the behaviour of a particle detector in the
$\alpha$-vacuum. One considers a monopole detector coupled to the
scalar field $\phi$ by the interaction
\be
g \int dt \phi (x(t)) m(t) ,
\ee
where $x(t)$ is the path followed by the particle detector,
with proper time $t$, and $m(t)$ is an operator acting on internal
states of the detector. If the operator $m(t)$ has eigenstates $|E_i
\ket$ with energies $E_i,$ we define the matrix elements $m_{ij}$ as 
\be 
m_{ij} = \bra E_i|m(0) |E_j \ket.
\ee
Then the probability that the detector
reports a change in energy from $E_i$ to $E_j,$ \mbox{ $ E_j >E_i,$} is
\be 
P(E_i \to E_j)= g^2 |m_{ij}|^2 \int_{-\infty}^{\infty} dt
dt' e^{-i(E_j -E_i)(t'-t) }G(x(t'), x(t) ), 
\ee   
where $G(x,x')$ is the Wightman function. Substituting in
(\ref{eq:galpha}),~\cite{burg,bms} found
\be
\frac{P(E_i \to E_j)}{P(E_j \to E_i)} = e^{-2\pi \Delta E} \left|
\frac{1 + e^{\alpha + \pi \Delta E}}{1 + e^{\alpha-\pi \Delta E}}
    \right|^2 
\label{dStrans}
\ee
showing that the detector has a non-thermal response. 

Note that in particular, at high energies the detector response
becomes independent of the energy difference. This is a sign of the
bad short-distance behaviour of the $\alpha$-vacua. From
(\ref{gafromge}), one can see that the short-distance
singularity of the $\alpha$-vacuum Wightman function is related to the
singularity in the Euclidean Wightman function by a factor of
$N_{\alpha}^2 (1 + e^{\alpha+\alpha*})$. Thus, as $x \to x'$, 
\be
G_{\alpha} (x, x') = N_{\alpha}^2 \left[ 1+e^{\alpha +\alpha ^*}
\right] 
 \frac{ 1 }{(2 \pi)^2 \sigma (x, x')} + \ldots,
\ee
where $\sigma (x,x')$ is half the square of the geodesic distance between
$x$ and $x'$, and 
$\ldots$ denotes less singular terms. The unusual coefficient of the
singularity implies that the $\alpha$-vacua are not Hadamard states:
they have a different short-distance singularity than the flat-space
vacuum propagator.

\subsection{Vacuum ambiguity in bubble and black hole}

Since the spacetimes we are interested in have a de Sitter factor,
they will naturally all have a similar vacuum ambiguity. That is, if
we consider a massive scalar field on either the bubble or the black
hole, we can again choose a mode expansion satisfying
\be
\phi_n^E(x_A)=\phi_n^{E*}(x),
\ee
where $x_A$ is the opposite point to $x$ only in respect to the
de Sitter factor coordinates.  If $x$ has full coordinates $\chi, r,
\tau, \theta $ and $\phi,$ then $x_A$ has coordinates $\chi, r, -\tau,
\theta +\pi$, and $ \phi +\pi $. Then we can define Bogoliubov transforms on
the Euclidean modes of this five-dimensional space in an identical way
to eq.\eqref{eq:mamodes}, to obtain new modes $\tilde{ \phi}_n (x)$
and similarly from the operators on this space, $a_n^E,$ we can define
$\tilde{a}_n$. As a consequence there are also multiple vacua in this
space, parametrised by $\alpha, $ satisfying $\tilde{a}_n |\alpha \ket
=0.$ The associated Wightman functions $G_{\alpha}$ defined using
these new vacua or mode expansions can again be expressed in terms of
the Euclidean Wightman function on the bubble, $G_E,$ by
\eqref{gafromge}.\footnote{For the locally AdS black hole, it is
possible to have such a vacuum ambiguity despite the fact that AdS has
a unique invariant vacuum state because the quotient broke the
$SO(2,4)$ isometry group to $SO(1,1) \times SO(1,3)$, which is no
longer sufficient to determine a unique vacuum state.} This
$\alpha$-vacuum ambiguity in the bulk is clearly related to a
corresponding $\alpha$-vacuum ambiguity in the boundary.

For the bubble of nothing solution (and of course for the boundary spacetime),
the de Sitter coordinates of (\ref{bubble}) cover the whole spacetime,
and the de Sitter factor does not degenerate anywhere. The physics of
these $\alpha$-vacua is hence little different from the familiar
discussion in de Sitter space, and we will not elaborate on it
here. The situation is more interesting for the locally AdS black
hole, however, so we pursue this case in more detail. 

First, we note that in the black hole background, we can find $G_E$
explicitly.  This is done by using the form of the propagator in full
AdS$_d$ \cite{burgess}. In AdS, the invariance under $SO(2,4)$ implies
that the propagator can only be a function of the geodesic distance
between the two points $x,x'$, $2\sqrt{\sigma(x,x')}$. It is in fact
convenient to express the propagator as a function of $P = X\cdot X'$,
where $X, X'$ are the corresponding points in the embedding
coordinates and the inner product is with respect to the metric of
signature $(--++++)$ in the embedding space. This is related to
geodesic distance through $\sigma(x,x') = -(P+1)$. In terms of $P$,
the propagator is the solution to
\be 
\frac{1}{ (P^2 -1 )^{ \frac{d-2}{2} }} \partial_P \left( ( P^2
-1)^{\frac{d}{2}} \partial_P G \right) - m^2 G =0 
\ee
which is regular at $P=\infty$. The required function $G$ is
\be
G_E(P)= \frac{-e^{ i\pi d /2 }\sqrt{\pi}  \Gamma(2b) }{2^{c-
\frac{1}{2} } \Gamma (c)} (P^{-2})^b \hspace{0.5em} _2F_1
\left(b+\frac{1}{2}  , b\hspace{0.5em} ; c \hspace{0.5em} ; P^{-2}
\right), 
\ee
where 
\be 
b= \frac{-1}{4} + \frac{1}{4}d + \frac{1}{4}\sqrt{ (d-1)^2 +4m^2 },\quad
c=1+\frac{1}{2}\sqrt{ (d-1)^2 +4m^2}.
\ee
It is divergent for $P=0$ and for all $|P|=1$ \cite{hyperbook}. Those
points which are light-like separated have $P=-1.$  

To write the propagator for the quotiented space, we must sum over the
images due to the periodic identification in the $\chi$ direction. The
Euclidean Wightman function on the locally AdS black hole can then be
expressed as
\be \label{geonbh}
G_E(P(x, x')) \sim \sum_{n=-\infty}^{\infty} P(x, {x'}_n)^{-2b}
\hspace{0.5em} _2F_1 \left( (b+\frac{1}{2}) , b \hspace{0.5em} ; c
\hspace{0.5em}; P(x, {x'}_n)^{-2} \right), 
\ee
where the $n$ dependence of $x'$ indicates that we have included every
image of $x'$ under the identification. 

This knowledge of the Euclidean Wightman function is sufficient to
determine the $\alpha$-vacua Wightman functions, by using the formula
\eqref{gafromge} relating it to an expression with four terms, each
involving the Euclidean propagator.  Recall that two of these depend
only on $x$ and $x',$ but two involve the points antipodal to the
original positions.

\section{Singularities in the $\alpha$-vacuum on the black hole}
\label{bhalpha}

\begin{figure}[htbp]
\begin{center}
\psfragscanon
\includegraphics[scale=0.4]{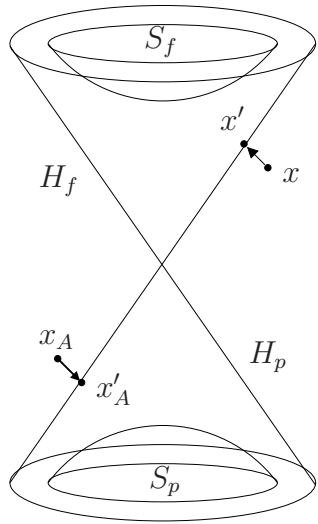}
\caption{As $x \to x'$ on the event horizon, extra singularities
  appear due to the lightlike separation of $x$ and $x'_A.$ \label{fig3}} 
\end{center}
\end{figure}

The important property of the $\alpha$-vacuum Wightman function for
the locally AdS black hole is that it develops new singularities on
(and inside) the event horizon. These arise because the antipodal
points, which were always separated by the cosmological horizon in de
Sitter space, become causally related. In the coordinates of
(\ref{bhk}), the antipode of a point $y^\mu, \chi$ is the point
$-y^\mu, \chi$, and on and inside the light cone \mbox{$y^2=0$,} these points
are causally separated. This means that on the horizon, there are
additional short-distance singularities in the propagator: in the
expression
\be G_{\alpha }(x, x')=N _{\alpha}^2 (G_E(x, x') +e^{\alpha +\alpha
^*}G_E(x', x) +e^{\alpha^*}G_E(x, {x'}_A) +e^{\alpha}G_E(x_A, x')), 
\ee 
the last two terms can produce an additional singularity as $x \to x'$
if $x=x'$ is a point on the event horizon, as shown in
figure~\ref{fig3}. Because of the antipodal map involved, this
additional singularity will have a completely different structure from
the usual short-distance singularity: it is not simply proportional to
$1/\sigma (x,x')$. It is thus potentially more dangerous than the
previous failure of the $\alpha$-vacuum to be Hadamard in de Sitter
space.

We will argue that these singularities are a sign that the
$\alpha$-vacua are unphysical on this black hole spacetime. We will
see below that they lead to a breakdown in the stress tensor on the
event horizon, and obstruct attempts to probe the region behind the
horizon by analytic continuation. 

They are also interesting from the point of view of understanding the
locally AdS black hole, as they are a clear sign that the event
horizon is a special place in the black hole geometry. We think of
these new singularities as analogous to the breakdown of a state with
the wrong temperature or a non-thermal state on the event horizon of a
Schwarzschild black hole, which selects the usual Hartle-Hawking state
as the unique regular vacuum~\cite{kaywald}.

\subsection{Particle detectors}

We now seek signs of these new singularities in the Wightman function
in physical observables. We will first examine whether a particle
detector crossing the horizon sees anything special, following the
calculation reviewed in section~\ref{alpharev}. As before, the
probability that the detector reports a change in energy from $E_i$ to
$E_j, E_j >E_i,$ is
\be \label{eq:pdr1}
P(E_i \to E_j)= g^2 |m_{ij}|^2 \int_{-\infty}^{\infty} dt
dt' e^{-i(E_j -E_i)(t'-t) }G_\alpha(x(t'), x(t) ). 
\ee 

We can consider different trajectories for the detector. If we take
the detector to stay at a constant $r$ in (\ref{bh}), staying outside
the black hole, then the behaviour in an $\alpha$-vacuum will be the
same as the behaviour in the de Sitter case, and the ratio of
transition probabilities will be given by (\ref{dStrans}). 

Consider instead an inertial observer, who freely falls across the
black hole horizon following some geodesic. Consider first the
Euclidean vacuum. Then since $G_E(t'-t)$ is holomorphic in the lower
half plane, we can close the contour of integration in \eqref{eq:pdr1}
in the LHP to find that the probability is zero. This is the expected
result, since the black hole solution is locally AdS, and the
Euclidean vacuum is obtained by sum over images from the usual AdS
vacuum. To calculate the detector response in the $\alpha$ vacuum, we
consider for definiteness a geodesic through the origin, so that
$x_A(t)=x(-t)$ {}\footnote{The geodesic used is not important, as the
argument essentially relies only on the fact that the Euclidean
propagator depends only on the difference in proper time along the
path, which is true for any geodesic in AdS.}. Then, using the
expression \eqref{gafromge} for the $\alpha$-vacuum propagator in
terms of the Euclidean propagator, the detector response
\eqref{eq:pdr1} is
\begin{multline}
P_{\alpha}(E_i \to E_j )= \int_{-\infty}^{\infty} dt dt' e^{-i(E_j -
E_i )(t'-t)} N_{\alpha}^2 \\ 
\times \left( G_E(t, t') + e^{\alpha +\alpha^*}G_E(t',
t)+e^{\alpha}G_E(-t, t') +e^{\alpha^*} G_E(t, -t') \right). 
\end{multline}

In the last two terms, where the Wightman function involved is a
function only of $t+t'$, we can immediately perform the integral over
$t-t'$ by closing the contour in the LHP to get $0$, showing that
these terms make no contribution. The first term is merely
proportional to the original integral in the Euclidean vacuum, so it
too gives no contribution.

The remaining term involves $G_E(t',t) = G_E(t'-t) =
G_E(-t,-t')$. Thus, we have 
\be
P_{\alpha}(E_i \to E_j )= \int_{-\infty}^{\infty} dt dt' e^{ -i (E_j -
E_i)(t' - t)} N_{\alpha}^2 e^{\alpha +\alpha^*} G_E( -t, -t'). 
\ee
Upon changing variables from $t, t'$ to $-t, -t',$ the exponential
picks up a minus sign, and we are left with the Euclidean rate for the
detector to see a change in energies from $E_j$ to $E_i,$ 
\be 
{P}_{\alpha}(E_i \to E_j )= N_{\alpha}^2 e^{\alpha +\alpha ^*}
P_E\left( E_j \to E_i \right).  
\ee
Similarly, if we consider the probability for the transition from $E_j
\to E_i $ in the $\alpha$-vacuum, the only contribution will come from
the first term in (\ref{gafromge}), giving\footnote{These rates are
divergent because of the integration over $t+t'$, but this simply
gives an overall factor that will cancel when considering the ratio of
rates.}
\begin{align}
P_{\alpha} (E_j \to E_i) &= \int_{-\infty}^{\infty} dt dt' e^{ -i (E_i
- E_j)(t' -t)} N_\alpha^2 G_{E}(x(t'), x(t))\\ 
&=N_\alpha^2 P_{E}(E_j \to E_i).
\end{align}
 The ratio of the two rates is
\be
\frac{ P_{\alpha} \left( E_i \to E_j \right) }{ P_{\alpha} \left( E_j
\to E_i \right) } = e^{\alpha +\alpha^*}. 
\ee
%

We note that this result is independent of the energies involved,
which might seem a disturbing result, but this is the usual problem
with the short-distance structure of the $\alpha$-vacuum,
corresponding precisely to the high energy behaviour of
(\ref{dStrans}). There is no sign in this calculation of the
additional singularities at the event horizon, because the relevant
parts of the $\alpha$-vacuum propagator made no contribution to the
calculation. This is perhaps surprising; if there is a breakdown in
the quantum state on the horizon, we would expect the behaviour of a
particle detector crossing the horizon to be affected. 

\subsection{Stress-energy tensor}

Another natural observable to consider in looking for reflections of
this new singularity in the propagator on the horizon is the
expectation value of the stress tensor in the $\alpha$-vacuum, $\bra
T_{\mu\nu} \ket_\alpha$. However, the usual construction of this
quantity relies on the assumption that the state is Hadamard
(see~\cite{waldr} for a review).  As we have already noted, the
$\alpha$-vacua are not Hadamard states. Hence, we cannot define a
stress tensor by the normal procedure in an $\alpha$-vacuum state on
any spacetime, and it therefore does not appear to be available to us
as a probe of the new singularities in the state on the horizon of the
locally AdS black hole.

Fortunately, Bernard and Folacci~\cite{bernard} have overcome this
obstacle and defined a renormalised expectation value for the stress
tensor in the $\alpha$-vacua of de Sitter space despite the non-Hadamard
form of the short-distance singularity.

To briefly review, we usually construct the stress tensor by taking
a coincidence limit of an appropriate differential operator acting on
the bi-distribution:
\be
\bra T_{\mu\nu}(x) \ket_\alpha = \lim_{x \to x'} {\cal D}_{\mu\nu}
F(x,x'),
\label{stresst}
\ee
where ${\cal D}_{\mu\nu}$ is a differential operator determined from
the Lagrangian whose precise form is not important for the present
purpose. We cannot take the bi-distribution $F(x,x')$ to be the
Wightman function, as its singularity as $x \to x'$ would produce a
divergent result for $\bra T_{\mu\nu}(x) \ket$. We must first
renormalize: this is done by defining
\be
F(x,x') = G(x,x') - H(x,x'),
\label{sub}
\ee
where $H(x,x')$ is the Hadamard bi-distribution. This  has the form 
\be
H (x, x') = \frac{1}{(2\pi)^2} \frac{  U(x, x') }{\sigma(x, x')}+ V(x,
x') \log \sigma (x, x') +W(x, x'),
\label{had}
\ee
where $\sigma$ is half the square of the geodesic distance between $x$
and $x'$, and $U(x,x)=1$.  In the massive case, the functions $U, V,
W$ can be determined by requiring that $H$ satisfy the Klein-Gordon
equation in each argument. Thus, this renormalisation is
state-independent, being entirely determined by the geometry of the
spacetime manifold. 


If $G(x,x')$ is of the Hadamard form, the subtraction in (\ref{sub})
will cancel the divergences in the two functions, allowing us to
define a finite renormalised stress tensor via
(\ref{stresst}). However, in the case of interest, the Wightman
function  for an $\alpha$-vacuum $G_\alpha(x,x')$ is not of Hadamard
form. This subtraction will not then define a good renormalised stress
tensor. 

The procedure adopted in~\cite{bernard} to address this problem was to
introduce a small element of state-dependence into the renormalisation
procedure: when we are considering an $\alpha$-vacuum in de Sitter
space, for which 
\be
\lim_{x \to x'} G_\alpha(x,x') = N_\alpha^2 e^{\alpha+\alpha^*}
\left( \frac{1}{(2\pi)^2} \frac{1}{\sigma(x, x')}+ \bar{V}(x,
x') \log \sigma (x, x') +\bar{W}(x, x') \right)
\ee
for some $\bar{V}$ and $\bar{W}$, then we define\footnote{There is a
difference in notation between our paper and~\cite{bernard}, so the
$\alpha$ used there is not the same as the one used here.}
\be
F(x,x') = G(x,x') - N_\alpha^2 e^{\alpha+\alpha^*} H(x,x'),
\label{newsub}
\ee
where $H(x,x')$ is given by (\ref{had}), and use (\ref{stresst}) for
this $F$ to define the stress tensor. One can also think of this as
defining $\bra T_{\mu\nu} \ket[G_\alpha] = N_\alpha^2
e^{\alpha+\alpha^*} \bra T_{\mu\nu} \ket[(N_\alpha^2
e^{\alpha+\alpha^*})^{-1} G_\alpha]$, where $\bra T_{\mu\nu}
\ket[(N_\alpha^2 e^{\alpha+\alpha^*})^{-1} G_\alpha]$ is defined by
the usual procedure. The point is that $\bar{G} = (N_\alpha^2
e^{\alpha+\alpha^*})^{-1} G_\alpha$ is of Hadamard form. Although this
procedure introduces state-dependence into the renormalisation
procedure, the stress tensor so defined shares all the good properties
of the usual stress tensor: the difference in energy for excited
states above an $\alpha$-vacuum will be given by the usual
point-splitting expression; the construction remains local; and
$\nabla^\mu \bra T_{\mu\nu}\ket = 0$ (since the additional factor
introduced is a constant). This allows us to discuss the renormalised
stress tensor in the $\alpha$-vacua on de Sitter space. 

We can apply this same prescription to obtain a notion of $\bra
T_{\mu\nu} \ket$ for our $\alpha$-vacuum states on the bubble of
nothing or locally AdS black hole spacetimes. For the bubble of
nothing, this gives a finite well-defined stress tensor
everywhere. For the black hole, however, this prescription breaks down
on the black hole horizon: the new singularities arising from the
terms involving $G_E(x,x'_A)$ and $G_E(x_A,x')$ in (\ref{gafromge})
are not cancelled by the subtraction (\ref{newsub}), so the stress
tensor becomes ill-defined on the horizon. The new singularities imply
that $\bar{G} = (N_\alpha^2 e^{\alpha+\alpha^*})^{-1} G_\alpha$ fails
to be of Hadamard form on the horizon. 

Thus, the ill-behavedness of the $\alpha$-vacua on the horizon of the
locally AdS black hole is signalled through the breakdown of the
procedure of~\cite{bernard} for defining a renormalised stress
tensor. 

\subsection{Analytic continuation}

Another way to see the relation between the event horizon and
$\alpha$-vacua is to consider the extension of the analytic
continuation argument of~\cite{mal1,kos} to this case. In~\cite{kos},
it was shown that an $n$-point correlation function in the CFT on the
two boundaries of the eternal BTZ black hole could be related either
to bulk interactions integrated over the region $r \geq r_+$ in the
bulk spacetime or over the region $r \geq 0$, including the region
behind the event horizon (with a different $i \epsilon$ prescription
for the bulk-boundary propagators). The argument used analyticity
properties of the $n$-point function, deforming the contour integral over $r
\geq r_+$ to an integral over the Euclidean spacetime by complexifying
the time coordinate, and then rotating back to the real Lorentzian
section in Kruskal coordinates.

Despite the somewhat different structure of the spacetime in the
locally AdS black hole, we can apply a similar argument here. If we
consider an integral which initially runs over the region $r \geq 0$
in the Lorentzian black hole solution
\be
ds^2 = (1+\frac{r^2}{l^2}) d\chi ^2 + (1+\frac{r^2}{l^2})^{-1}dr^2
+r^2[-d\tau ^2 +\cosh^2\tau (d\theta ^2 + \sin ^2 \theta d\phi^2)],
\ee
we can continue to a Euclidean solution by $\tau \to -i\vartheta$, giving
\be
ds^2 = (1+\frac{r^2}{l^2}) d\chi ^2 + (1+\frac{r^2}{l^2})^{-1}dr^2
+r^2[d\vartheta^2 +\cos^2 \vartheta (d\theta ^2 + \sin ^2 \theta d\phi^2)].
\ee
We then define a new radial coordinate by $r = 2 l y/(1-y^2)$,
so 
\be
ds^2 =   {4 l^2 \over (1-y^2)^2}\left\{ dy^2  +y^2[d\vartheta^2 +\cos^2
    \vartheta (d\theta ^2 + \sin ^2 \theta d\phi^2)] \right\} +
{(1+y^2)^2 \over (1-y^2)^2} d\chi^2.  
\ee
We can recover the Kruskal coordinates of (\ref{bhk}) by defining
Cartesian coordinates $y^i$ $i = 1,\ldots,4$ on the
$(y,\vartheta,\theta,\phi)$ space, and analytically continuing $y^4
\to iy^0$. As in the BTZ case, the Cartesian coordinates are
restricted to the interior of the unit ball $(y^i)^2 < 1$; however,
the exterior of the unit ball is isometric to the interior, so we can
take the integral to run over all $y^i$ if we divide by a
factor of two. The factor of two is then used to convert the integral
over all $y^\mu$ in the Kruskal coordinates (\ref{bhk}) to an integral
over $-1 < y^\mu y_\mu <1$, covering the full black hole spacetime.

The above procedure is possible only in the Euclidean vacuum. If we
consider an $\alpha$-vacuum on the black hole, it will not be possible
to continue the integral in this way, as the $\alpha$-vacuum does not
define a regular propagator on the Euclidean spacetime when we
analytically continue $\tau \to -i\vartheta$. That is, the additional
pole associated with the extra divergences on the event horizon in an
$\alpha$-vacuum will obstruct this kind of contour deformation
argument. Thus, we see again that the $\alpha$-vacuum runs into
trouble when we try to look inside the black hole.

\section{Conclusions}
\label{concl}

The relation between the bulk and boundary for these time-dependent
spacetimes has several new and interesting features. The
identification of the cosmological/acceleration horizons in the bulk
with the de Sitter cosmological horizon in the boundary provides an
example of a new way of relating thermodynamic behaviour in the bulk
to the boundary. We believe this kind of relation should be very
general, applying to any non-compact horizon encountered in the
AdS/CFT correspondence. This identification provides new insight into
the thermodynamic interpretation of horizons in spacetime, since the
boundary interpretation for our bulk horizons strongly suggests that
the relation $S = A/4G$ can be applied even to such non-compact
horizons. This provides support for the very general connection
between entropy and horizons advocated in~\cite{jac,jpar}. We have
also argued that the area of the event horizon in the locally AdS
black hole should not be interpreted thermodynamically. We argued this
from a purely spacetime point of view, but it also seems a natural
result from the CFT side, since we conjectured that the CFT dual is a
vacuum state, where we see no role for an increasing entropy.

A central result of our paper was to show that the analogues of
$\alpha$-vacua for a free scalar field on the locally AdS black hole
spacetime break down on the event horizon. Thus, these are not good
quantum states on the full black hole solution. The unique regular
invariant vacuum state on the locally AdS black hole is the Euclidean
vacuum. We regard this as evidence that there are no $\alpha$-vacua in
the strongly-coupled dual field theory, which lives on de Sitter space
cross a circle. This provides support, from a very different
perspective, for the view taken by some authors that $\alpha$-vacua
are not good states in an interacting field theory from the point of
view of perturbation theory~\cite{banks,el1,el2,gold2}.
This selection of the Euclidean vacuum as a preferred state provides
an interesting example of how the bulk spacetime picture can be used
to study issues of quantum field theory on more general backgrounds.

It would be interesting to investigate further the interpretation of
these two spacetimes from the boundary point of view. In particular,
the fact that the bubble of nothing solution exists only if the radius
of the $\chi$ circle is less than a maximum value seems quite
mysterious from the boundary point of view. 

In fact, there is also another issue of interpretation which remains
open at a purely spacetime level. In the asymptotically flat case, the
bubble of nothing solution is interpreted as describing a
non-perturbative instability of the Kaluza-Klein vacuum, that is, flat
space with one spatial direction periodically
identified~\cite{wittenkk}. In the case of a negative cosmological
constant, the analogous interpretation of the bubble solution would be
to regard it as describing a non-perturbative decay of this quotient
of AdS, the locally AdS black hole. However, the fact that this
background is itself time dependent (and even has an event horizon!)
may complicate this interpretation.  See~\cite{ptoapp} for further
discussion of the interpretation of this bubble as describing a
non-perturbative instability.

\medskip
\centerline{\bf Acknowledgements}
\medskip    

We thank Vijay Balasubramanian and Joan Simon for useful
discussions. This work is supported by the EPSRC.

\bibliographystyle{utphys}  
 
\bibliography{alphainbubble2}   

\providecommand{\href}[2]{#2}\begingroup\raggedright\begin{thebibliography}{10}

\bibitem{maldads}
J.~M. Maldacena, ``The large {N} limit of superconformal field theories and
  supergravity,'' Adv. Theor. Math. Phys. {\bf 2} (1998) 231--252,
\href{http://xxx.lanl.gov/abs/hep-th/9711200}{{\tt hep-th/9711200}}.

\bibitem{ah}
O.~Aharony, S.~S. Gubser, J.~M. Maldacena, H.~Ooguri, and Y.~Oz, ``Large {N}
  field theories, string theory and gravity,'' Phys. Rept. {\bf 323} (2000)
  183--386,
\href{http://xxx.lanl.gov/abs/hep-th/9905111}{{\tt hep-th/9905111}}.

\bibitem{birmingham}
D.~Birmingham and M.~Rinaldi, ``Bubbles in anti-de {S}itter space,'' Phys.
  Lett. {\bf B544} (2002) 316--320,
\href{http://xxx.lanl.gov/abs/hep-th/0205246}{{\tt hep-th/0205246}}.

\bibitem{ross1}
V.~Balasubramanian and S.~F. Ross, ``The dual of nothing,'' Phys. Rev. D {\bf
  66} (2002) 086002,
\href{http://xxx.lanl.gov/abs/hep-th/0205290}{{\tt hep-th/0205290}}.

\bibitem{eva}
O.~Aharony, M.~Fabinger, G.~T. Horowitz, and E.~Silverstein, ``Clean
  time-dependent string backgrounds from bubble baths,'' JHEP {\bf 07} (2002)
  007,
\href{http://xxx.lanl.gov/abs/hep-th/0204158}{{\tt hep-th/0204158}}.

\bibitem{wittenkk}
E.~Witten, ``Instability of the {K}aluza-{K}lein vacuum,'' Nucl. Phys. {\bf
  B195} (1982)
481.

\bibitem{cvetic}
M.~Cvetic, S.~Nojiri, and S.~D. Odintsov, ``Cosmological anti-desitter
  space-times and time-dependent ads/cft correspondence,'' Phys. Rev. {\bf D69}
  (2004) 023513,
\href{http://xxx.lanl.gov/abs/hep-th/0306031}{{\tt hep-th/0306031}}.

\bibitem{ban}
M.~Banados, ``Constant curvature black holes,'' Phys. Rev. D {\bf 57} (1998)
  1068--1072,
\href{http://xxx.lanl.gov/abs/gr-qc/9703040}{{\tt gr-qc/9703040}}.

\bibitem{bgm}
M.~Banados, A.~Gomberoff, and C.~Martinez, ``Anti-de {S}itter space and black
  holes,'' Class. Quant. Grav. {\bf 15} (1998) 3575--3598,
\href{http://xxx.lanl.gov/abs/hep-th/9805087}{{\tt hep-th/9805087}}.

\bibitem{btz}
M.~Banados, C.~Teitelboim, and J.~Zanelli, ``The black hole in
  three-dimensional space-time,'' Phys. Rev. Lett. {\bf 69} (1992) 1849--1851,
\href{http://xxx.lanl.gov/abs/hep-th/9204099}{{\tt hep-th/9204099}}.

\bibitem{bhtz}
M.~Banados, M.~Henneaux, C.~Teitelboim, and J.~Zanelli, ``Geometry of the (2+1)
  black hole,'' Phys. Rev. D {\bf 48} (1993) 1506--1525,
\href{http://xxx.lanl.gov/abs/gr-qc/9302012}{{\tt gr-qc/9302012}}.

\bibitem{chernikov}
N.~A. Chernikov and E.~A. Tagirov, ``Quantum theory of scalar fields in de
  {S}itter space-time,'' Annales Poincare Phys. Theor. {\bf A9} (1968)
109.

\bibitem{tagirov}
E.~A. Tagirov, ``Consequences of field quantization in de {S}itter type
  cosmological models,'' Ann. Phys. {\bf 76} (1973)
561--579.

\bibitem{mot}
E.~Mottola, ``Particle creation in de {S}itter space,'' Phys. Rev. D {\bf 31}
  (1985)
754.

\bibitem{all}
B.~Allen, ``Vacuum states in de {S}itter space,'' Phys. Rev. D {\bf 32} (1985)
3136.

\bibitem{aj}
B.~Allen and T.~Jacobson, ``Vector two point functions in maximally symmetric
  spaces,'' Commun. Math. Phys. {\bf 103} (1986)
669.

\bibitem{dr}
I.~T. Drummond, ``Dimensional regularization of massless theories in spherical
  space-time,'' Nucl. Phys. {\bf B94} (1975)
115.

\bibitem{gh}
G.~W. Gibbons and S.~W. Hawking, ``Cosmological event horizons, thermodynamics,
  and particle creation,'' Phys. Rev. D {\bf 15} (1977)
2738--2751.

\bibitem{bros1}
J.~Bros and U.~Moschella, ``Two-point functions and quantum fields in de
  {S}itter universe,'' Rev. Math. Phys. {\bf 8} (1996) 327--392,
\href{http://xxx.lanl.gov/abs/gr-qc/9511019}{{\tt gr-qc/9511019}}.

\bibitem{bros2}
J.~Bros, H.~Epstein, and U.~Moschella, ``Analyticity properties and thermal
  effects for general quantum field theory on de {S}itter space-time,'' Commun.
  Math. Phys. {\bf 196} (1998) 535--570,
\href{http://xxx.lanl.gov/abs/gr-qc/9801099}{{\tt gr-qc/9801099}}.

\bibitem{banks}
T.~Banks and L.~Mannelli, ``De {S}itter vacua, renormalization and locality,''
  Phys. Rev. D {\bf 67} (2003) 065009,
\href{http://xxx.lanl.gov/abs/hep-th/0209113}{{\tt hep-th/0209113}}.

\bibitem{el1}
M.~B. Einhorn and F.~Larsen, ``Interacting quantum field theory in de {S}itter
  vacua,'' Phys. Rev. D {\bf 67} (2003) 024001,
\href{http://xxx.lanl.gov/abs/hep-th/0209159}{{\tt hep-th/0209159}}.

\bibitem{uhd1}
U.~H. Danielsson, ``On the consistency of de {S}itter vacua,'' JHEP {\bf 12}
  (2002) 025,
\href{http://xxx.lanl.gov/abs/hep-th/0210058}{{\tt hep-th/0210058}}.

\bibitem{gl1}
K.~Goldstein and D.~A. Lowe, ``A note on alpha-vacua and interacting field
  theory in de {S}itter space,'' Nucl. Phys. {\bf B669} (2003) 325--340,
\href{http://xxx.lanl.gov/abs/hep-th/0302050}{{\tt hep-th/0302050}}.

\bibitem{el2}
M.~B. Einhorn and F.~Larsen, ``Squeezed states in the de {S}itter vacuum,''
  Phys. Rev. D {\bf 68} (2003) 064002,
\href{http://xxx.lanl.gov/abs/hep-th/0305056}{{\tt hep-th/0305056}}.

\bibitem{collins1}
H.~Collins, R.~Holman, and M.~R. Martin, ``The fate of the alpha-vacuum,''
  Phys. Rev. D {\bf 68} (2003) 124012,
\href{http://xxx.lanl.gov/abs/hep-th/0306028}{{\tt hep-th/0306028}}.

\bibitem{gold2}
K.~Goldstein and D.~A. Lowe, ``Real-time perturbation theory in de {S}itter
  space,'' Phys. Rev. D {\bf 69} (2004) 023507,
\href{http://xxx.lanl.gov/abs/hep-th/0308135}{{\tt hep-th/0308135}}.

\bibitem{collins2}
H.~Collins and R.~Holman, ``Taming the alpha vacuum,''
\href{http://xxx.lanl.gov/abs/hep-th/0312143}{{\tt hep-th/0312143}}.

\bibitem{dBJM}
J.~de~Boer, V.~Jejjala, and D.~Minic, ``Alpha-states in de {S}itter space,''
\href{http://xxx.lanl.gov/abs/hep-th/0406217}{{\tt hep-th/0406217}}.

\bibitem{collins3}
H.~Collins, ``Fermionic alpha-vacua,''
\href{http://xxx.lanl.gov/abs/hep-th/0410229}{{\tt hep-th/0410229}}.

\bibitem{bernard}
D.~Bernard and A.~Folacci, ``Hadamard function, stress tensor and de {S}itter
  space,'' Phys. Rev. D {\bf 34} (1986)
2286.

\bibitem{mal1}
J.~M. Maldacena, ``Eternal black holes in {Anti-de-Sitter},'' JHEP {\bf 04}
  (2003) 021,
\href{http://xxx.lanl.gov/abs/hep-th/0106112}{{\tt hep-th/0106112}}.

\bibitem{kos}
P.~Kraus, H.~Ooguri, and S.~Shenker, ``Inside the horizon with {AdS/CFT},''
  Phys. Rev. D {\bf 67} (2003) 124022,
\href{http://xxx.lanl.gov/abs/hep-th/0212277}{{\tt hep-th/0212277}}.

\bibitem{cai}
R.-G. Cai, ``Constant curvature black hole and dual field theory,'' Phys. Lett.
  {\bf B544} (2002) 176--182,
\href{http://xxx.lanl.gov/abs/hep-th/0206223}{{\tt hep-th/0206223}}.

\bibitem{ross2}
J.~Figueroa-O'Farrill, O.~Madden, S.~F. Ross, and J.~Simon, ``Quotients of
  {AdS$_{p+1} \times S^q$}: Causally well-behaved spaces and black holes,''
  Phys. Rev. D {\bf 69} (2004) 124026,
\href{http://xxx.lanl.gov/abs/hep-th/0402094}{{\tt hep-th/0402094}}.

\bibitem{bk}
V.~Balasubramanian and P.~Kraus, ``A stress tensor for anti-de {S}itter
  gravity,'' Commun. Math. Phys. {\bf 208} (1999) 413--428,
\href{http://xxx.lanl.gov/abs/hep-th/9902121}{{\tt hep-th/9902121}}.

\bibitem{hs}
M.~Henningson and K.~Skenderis, ``The holographic {W}eyl anomaly,'' JHEP {\bf
  07} (1998) 023,
\href{http://xxx.lanl.gov/abs/hep-th/9806087}{{\tt hep-th/9806087}}.

\bibitem{skend}
K.~Skenderis, ``Asymptotically anti-de {S}itter spacetimes and their stress
  energy tensor,'' Int. J. Mod. Phys. {\bf A16} (2001) 740--749,
\href{http://xxx.lanl.gov/abs/hep-th/0010138}{{\tt hep-th/0010138}}.

\bibitem{strom}
A.~Strominger, ``Black hole entropy from near-horizon microstates,'' JHEP {\bf
  02} (1998) 009,
\href{http://xxx.lanl.gov/abs/hep-th/9712251}{{\tt hep-th/9712251}}.

\bibitem{wittenbh}
E.~Witten, ``Anti-de {S}itter space, thermal phase transition, and confinement
  in gauge theories,'' Adv. Theor. Math. Phys. {\bf 2} (1998) 505--532,
\href{http://xxx.lanl.gov/abs/hep-th/9803131}{{\tt hep-th/9803131}}.

\bibitem{kaywald}
B.~S. Kay and R.~M. Wald, ``Theorems on the uniqueness and thermal properties
  of stationary, nonsingular, quasifree states on space-times with a bifurcate
  {K}illing horizon,'' Phys. Rept. {\bf 207} (1991)
49--136.

\bibitem{jac}
T.~Jacobson, ``Thermodynamics of space-time: The {E}instein equation of
  state,'' Phys. Rev. Lett. {\bf 75} (1995) 1260--1263,
\href{http://xxx.lanl.gov/abs/gr-qc/9504004}{{\tt gr-qc/9504004}}.

\bibitem{jpar}
T.~Jacobson and R.~Parentani, ``Horizon entropy,'' Found. Phys. {\bf 33} (2003)
  323--348,
\href{http://xxx.lanl.gov/abs/gr-qc/0302099}{{\tt gr-qc/0302099}}.

\bibitem{bms}
R.~Bousso, A.~Maloney, and A.~Strominger, ``Conformal vacua and entropy in de
  {S}itter space,'' Phys. Rev. D {\bf 65} (2002) 104039,
\href{http://xxx.lanl.gov/abs/hep-th/0112218}{{\tt hep-th/0112218}}.

\bibitem{burg}
C.~J. Burges, ``The de {S}itter vacuum,'' Nucl. Phys. {\bf B247} (1984) 533.

\bibitem{burgess}
C.~P. Burgess and C.~A. Lutken, ``Propagators and effective potentials in
  anti-de {S}itter space,'' Phys. Lett. {\bf B153} (1985)
137.

\bibitem{hyperbook}
W.~Magnus and F.~Oberhettinger, {\em Special Functions of Mathematical
  Physics}.
\newblock Chelsea Publishing Company, 1949.
\newblock p.7.

\bibitem{waldr}
R.~M. Wald, {\em Quantum field theory in curved space-time and black hole
  thermodynamics}.
\newblock Chicago University Press, Chicago, USA, 1994.

\bibitem{ptoapp}
V.~Balasubramanian, K.~Larjo, and J.~Simon, ``Much ado about nothing,''. (to
  appear).

\end{thebibliography}\endgroup

\end{document}